# Refractive Index Engineering: Insights from Biological Systems for Advanced Optical Design

*Jiale Huang[1], Zihao Ou[2]\**

[1]Department of Biomedical Engineering, Johns Hopkins University, Baltimore, MD, 21218

[2]Department of Physics, The University of Texas at Dallas, Richardson, TX, 75080

ABSTRACT

This review explores the innovative design to achieve advanced optical functions in natural materials and intricate optical systems inspired by the unique refractive index profiles found in nature. By understanding the physical principles behind biological structures, we can develop materials with tailored optical properties that mimic these natural systems. One key area discussed is biomimetic materials design, where biological systems such as apple skin and the vision system inspire new materials. Another focus is on intricate optical systems based on refractive index contrast. These principles can be extended to design devices like waveguides, photonic crystals, and metamaterials, which manipulate light in novel ways. Additionally, the review covers optical scattering engineering, which is crucial for biomedical imaging. By adjusting the real and imaginary parts of the refractive index, we can control how much light is scattered and absorbed by tissues. This is particularly important for techniques like optical coherence tomography and multiphoton microscopy, where tailored scattering properties can improve imaging depth and resolution. The review also discusses various techniques for measuring the refractive index of biological tissues which provide comprehensive insights into the optical properties of biological materials, facilitating the development of advanced biomimetic designs. In conclusion, the manipulation of refractive index profiles in biological systems offers exciting opportunities for technological advancements. By drawing inspiration from nature and understanding the underlying physical principles, we can create materials and devices with enhanced performance and new functionalities. Future research should focus on further exploring these principles and translating them into practical applications to address real-world challenges.

**MAIN TEXT**

## 1. INTRODUCTION

The optical properties of biological systems are crucial for various functions in living organisms, such as environmental sensing and protection against harmful radiation.[1-4] While not all species rely heavily on vision, the eye is a significant optical units in many organisms.[5, 6] For example, in humans, light enters the eye through the cornea and lens, which focus the light to form an image on the retina. This process transmits and compresses information from the environment to the brain, enabling vision.[7] These complex optical functions are achieved through the synergistic cooperation of different cell groups and tissue types in the eye, each possessing unique optical properties and dynamic tunability.[8, 9] Understanding these principles has led to significant advancements in artificial optical systems and biomimetic designs.[10-13] For instance, researchers have studied the optical responses of cephalopods, which can change their skin color and texture for camouflage and communication.[14, 15] By mimicking these biological strategies, scientists have developed materials that can dynamically change color, transparency, and chemical sensing capabilities.[14, 16-18] These biomimetic materials hold great potential for applications in various fields, including adaptive camouflage, smart windows, and responsive sensors.[19, 20]

The optical functions of tissues are closely related to their refractive indices and subcellular structures.[21-23] Research on the refractive index of cells has evolved from measuring an average value to illustrating the heterogeneity of different subcellular components.[24] These complex refractive index profiles enable tissues to perform functions, such as guiding, filtering, focusing, reflecting, or scattering light (**Fig. 1**). For example, active optical tuning of the lens is achieved by changes in the refractive index gradient. Lens fiber cells the ciliary muscles adjust the lens shape to modify light refraction and focus on the eyes.[5, 9] Müller cells in the retina function like optical fibers, transmitting light with minimal dispersion to the photoreceptor layer.[25] Birds' photoreceptors contain oil droplets that grant them superior vision compared to humans. These oil droplets in cone photoreceptors act as spherical microlenses, filtering light to enhance color discrimination and constancy.[26] The skin of an apple contains organelles with pigments like chlorophylls and carotenoids, which change dynamically during fruit development. These organelles scatter specific wavelengths, imparting various colors.[27] Similarly, such strategies are also utilized in many passerine birds have feathers with melanin-based coloration.[28]

Structural coloring is a strategy that uses refractive index design inside biological systems.[10] Many species, including insects, birds, and fishes, all utilize this strategy to achieve the selective color coding they want.[29-32] When features have size similar to the wavelengths of light, the interactions between the light and features are strongly correlated with the size and structure. This forms a photonic crystal with periodic structures in refractive index at wavelength scale.[33] By mimicking these designs, artificial materials with hierarchical structures can be engineered to show functions that single optical components cannot achieve, such as selective enhancing light reflection without bleaching over time.[13] However, although the structural colors have been extensively discussed and utilized in the past research, the physical origin of the high refractive index materials in the biological tissue are usually overlooked. There are also many applications of leveraging the unique refractive index profiles in living biological tissue to achieve advanced optical functions, such as selective reflection and transmittance,[31] optical filtering,[26] optical focusing,[9] and waveguiding.[25]

In this review, inspired by intricate biological structures, we aim to summarize key engineering strategies for modulating refractive index within biological systems, distinct from structural coloring. We first explore the theoretical framework governing light interaction with biological tissues, emphasizing the physical definitions of refractive index and related optical characteristics of tissue. Next, we examine biological systems that exhibit refractive index variations to achieve distinct optical functionalities, including optical focusing, waveguiding, filtering, and coating, with examples from cellular structures to complex tissues. We then discuss methodologies and instruments for quantifying refractive index, highlighting their applications, especially for biological tissues. Finally, we discuss potential innovations based on the fundamental understanding of refractive index in biomedical imaging, diagnostics, and therapeutic technologies, offering insights into future research directions and practical implementations.

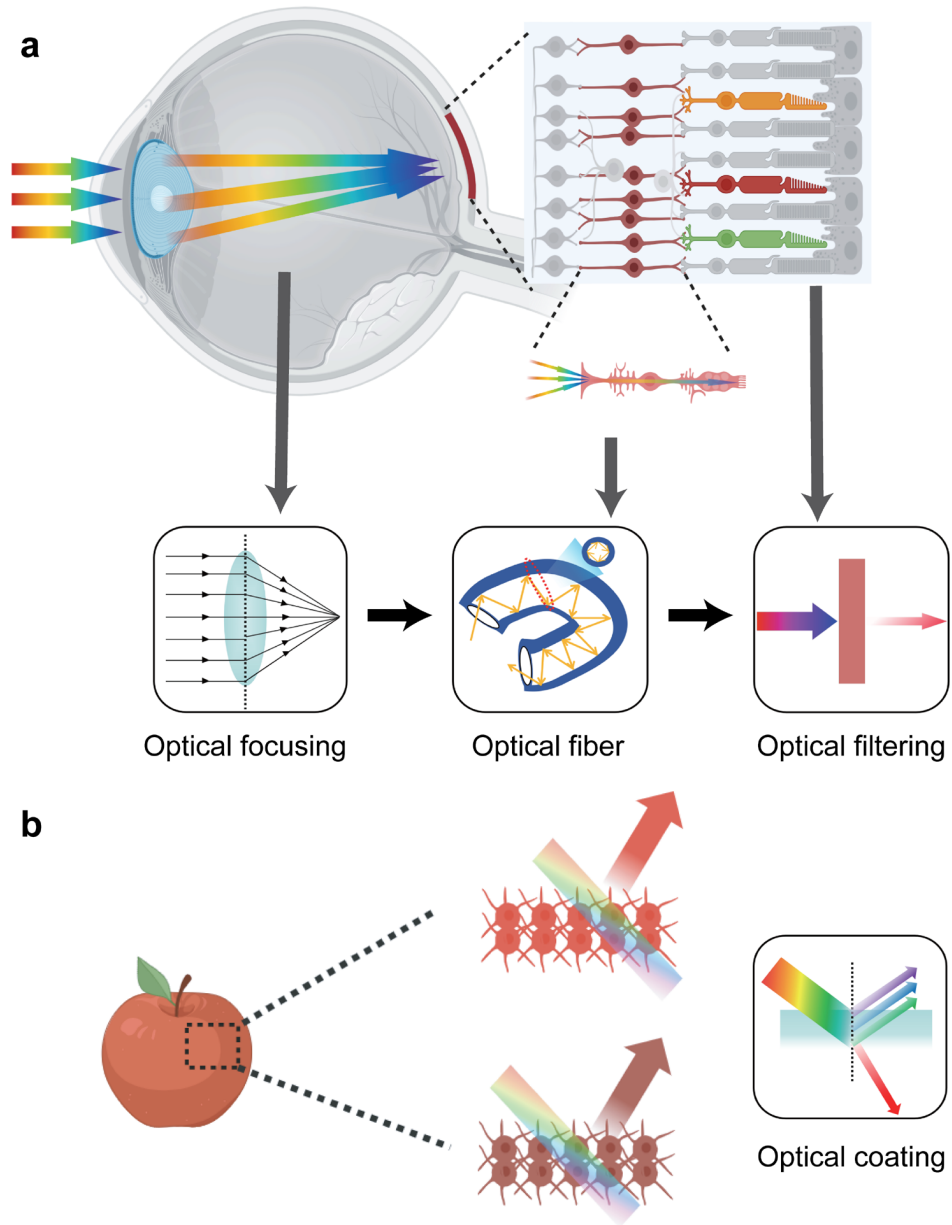

**Figure 1**. Refractive index engineering in the systems to achieve different optical functionalities, including (**a**) optical focusing, waveguide, filter in the vision system, and (**b**) optical coating in the plant.

## 2. REFRACTIVE INDEX AND OPTICAL PROPERTIES OF TISSUES

### 2.1. Physical Meaning of Refractive Index

The refractive index ($n$) is a key optical parameter that describes how light interacts with a medium.[34] To understand its physical meaning, we can consider the propagation of light with a fixed wavelength ($\lambda$), frequency ($\nu$), and angular frequency ($\omega = 2\pi\nu$). By solving Maxwell's equations in a uniform medium with a complex dielectric constant ($\varepsilon$), the plane wave solution for the electric field ($\boldsymbol{E}$) can be expressed as: $\boldsymbol{E}(\boldsymbol{r}, t) = \boldsymbol{E_0}e^{(i\boldsymbol{k}\cdot\boldsymbol{r} - i\omega t)}$. Here, $\boldsymbol{r}$ is the spatial coordinate, $t$ is time, and $\boldsymbol{E_0}$ is a fixed electric field vector indicating the polarization and magnitude of the electric field. The wave vector $\boldsymbol{k}$ indicates the direction of light propagation and has a magnitude of $2\pi n\lambda^{-1}$. Here, $n$ describes the interaction between the electromagnetic field and the material, including both the speed of light within the material and the attenuation of the wave during propagation. Mathematically, refractive index is the square root of dielectric constant ($n = \sqrt{\varepsilon}$). Similar to the dielectric constant, which has both real and imaginary components, the refractive index is a complex number ($n = n' + in''$), consisting of a real component $n'$ and an imaginary component $n''$.

### 2.1.1. Real Component of Refractive Index

The real component ($n'$) describes the phase velocity of light within the medium. In a vacuum, electromagnetic waves propagate at the speed of light ($c$). In other media with a refractive index $n$, the phase speed of light waves changes to $cn'^{-1}$. The direction of electromagnetic wave propagation also changes when transitioning from one medium to another (**Fig. 2a**). When light moves from one material to another, if $\theta_i$ represents the angle of incidence in the first material, and $\theta_r$ denotes the angle of refraction in the second medium, these angles are connected by Snell's law[34]: $n_i'\sin(\theta_i) = n_r'\sin(\theta_r)$. In a special case when $n_r' < n_i'$, $\theta_r$ will be larger than $\theta_i$. When $\theta_i \geq \theta_{ic} = \sin^{-1}\left(\frac{n_r'}{n_i'}\right)$, light will be reflected back to the first medium, known as total internal reflection, which is the principles behind refractometer discussed in Section 4. It is worth highlighting that the strong scattering of biological tissue originates from the heterogeneity of refractive index profiles in the tissue, which has the same physical origin as the refraction of light discussed above.[35-39]

### 2.1.2. Imaginary Component of Refractive Index

The imaginary component $n''$ represents the absorption of light by the medium during propagation. In a uniform medium with a fixed $n''$, the magnitude of the electric field decays exponentially $\boldsymbol{E}(\boldsymbol{r}, t) = \boldsymbol{E_0} \mathrm{e}^{(\mathrm{i}n'\boldsymbol{k_0} \cdot \boldsymbol{r} - \mathrm{i}\omega t)} \mathrm{e}^{-n''\boldsymbol{k_0} \cdot \boldsymbol{r}}$. Here $\boldsymbol{k_0}$ indicates the wave vector of light if propagating in vacuum. This exponential decay forms the basis of the well-known Lambert-Beer law, and the absorption coefficient $\mu_\mathrm{a}$ can be derived as $\mu_\mathrm{a} = 2|\boldsymbol{k_0}|n''$. Due to their different effects on light propagation, the real component $n'$ and the imaginary component $n''$ are often discussed separately in literature. Since both quantities describe the physical interaction between electromagnetic waves and materials, they are not physically independent and are mathematically related by the Kramers-Kronig relations.[40]

### 2.2. Kramers-Kronig Relations in Refractive Index

The real and imaginary parts of the refractive index are connected by the Kramers-Kronig relations. These relations are based on the principle of causality, which states that an effect cannot precede its cause.[40] In optics, causality imposes the limitation that no scattered wave can exist before the incident wave reaches the scattering center. Using complex analysis, the connection between the real and imaginary components can be described in the frequency domain as:

$$n'(\omega) - 1 = \frac{2}{\pi}\text{P.V.}\int_0^\infty \frac{\omega'n''(\omega')}{\omega'^2 - \omega^2}d\omega'$$

The connection between the imaginary and real parts of the refractive index can also be expressed in the wavelength domain: [41]

$$n'(\lambda) - 1 = \frac{2}{\pi}\text{P.V.}\int_0^\infty \frac{n''(\lambda')}{\lambda'\left(1 - \frac{\lambda'^2}{\lambda^2}\right)}d\lambda'$$

As illustrated in **Fig. 2b**, an arbitrary imaginary refractive index peak is introduced, and the modulation in the real part of the refractive index can be quantitatively evaluated. Combined with experimental methods that measure the imaginary part of the refractive index $n''$, this numerical tool offers a unique strategy to evaluate the real part of the refractive index $n'$ and we will discuss more details about these strategies in Section 4.

### 2.3. Scattering and Absorption in Biological Tissue

Different tissues have varying scattering and absorption properties, which can be compared in several published databases.[2, 42-44] The scattering properties of tissue are determined by its subcellular composition, including the anisotropic factor indicates whether light is scattered uniformly or in specific directions.[45] For most tissues, such as muscle and skin, scattering is 10-100 times greater than absorption, making it the main limiting factor in optical signal transportation. However, in tissues like blood, strong absorption by hemoglobin limits light penetration, especially near hemoglobin's absorption peaks.[45, 46]

### 2.3.1. Scattering Coefficient in Biological Tissue

Scattering within tissue is caused by the refractive index contrast between different components.[35-39] Understanding the subcellular refractive index contrast in various biological tissues is challenging, but several articles summarize research measuring such refractive index profiles.[23, 24, 47, 48] Instead of investigating the light-matter interaction at cellular level, bulk optical properties are used to quantify the absorption and scattering efficiency of tissue (**Fig. 2c**). For a scattering-only sample, the decay of non-scattered light $I$ can be described as: $I(d) = I_0 e^{-\mu_s d}$, where $I_0$ is the initial light intensity, $\mu_s$ is the scattering coefficient, and $d$ is the thickness of the sample. Apart from the strength of scattering inside the tissue, a dimensionless anisotropy factor $g$ is usually introduced to describe the level of isotropy after scattering. The scattering behaviors are related to the microscopic origins of light interaction with subcellular features, such as fibrils and organelles inside the cells.[45] In applications where the phase information of the light is not crucial, the total light intensity $I(d) = I_0 e^{-\mu_s' d}$ and $\mu_s'$ is the reduced scattering coefficient, defined by $\mu_s' = (1 - g)\mu_s$.[49]

### 2.3.2. Absorption Coefficient in Biological Tissue

Absorption in tissue is caused by molecules that absorb light at specific wavelengths, such as hemoglobin and melanin.[45, 50] **Table 1** provides examples of chemical structures of absorbing biomolecules. Absorption signatures can be used to measure the concentration of certain molecules within tissue. To quantify the absorption efficiency of tissue, an absorption coefficient $\mu_a$ is introduced, and the decay of light intensity in an absorbing-only medium can be described as: $I(d) = I_0 e^{-\mu_a d}$.

**Table 1.** Common absorbing molecules in biological systems.

| Chemical Name | Chemical Formula | Chemical Structure |
|---|---|---|
| astaxanthin | $C_{40}H_{52}O_4$ | 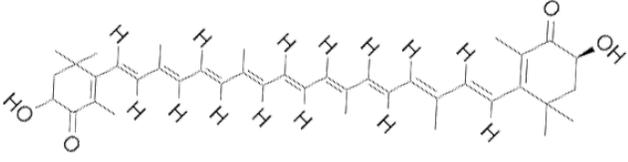 |
| anthocyanin | $C_{15}H_{11}O^+$ | 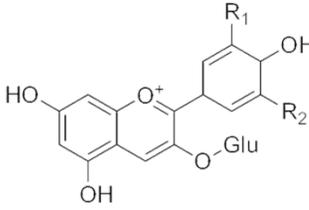 |
| galloxanthin | $C_{27}H_{38}O_2$ | 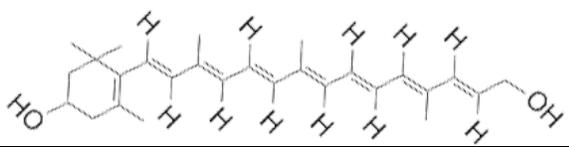 |
| ε-carotene | $C_{40}H_{56}$ | 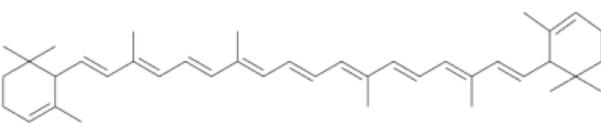 |
| β-carotene | $C_{40}H_{56}$ | 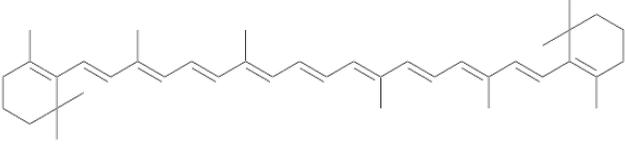 |
| chlorophyll-b | $C_{55}H_{70}MgN_4O_6$ | 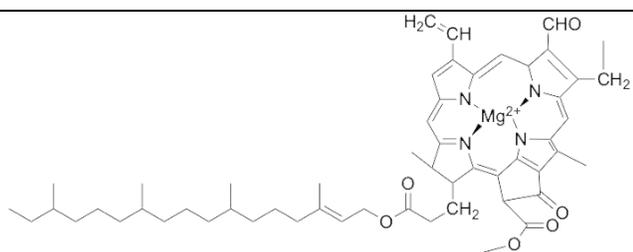 |
| chlorophyll-a | $C_{55}H_{72}MgN_4O_5$ | 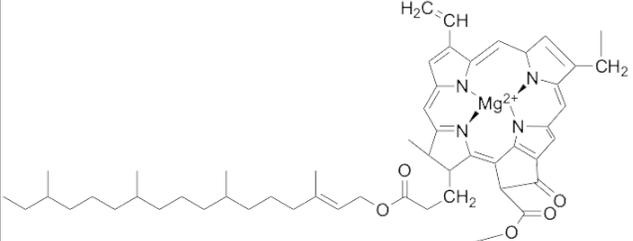 |
| ideain | $C_{21}H_{21}ClO_{11}$ | 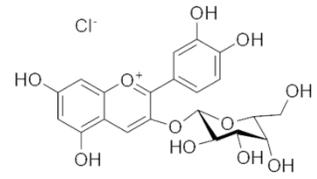 |
| hyperoside | $C_{21}H_{20}O_{12}$ | 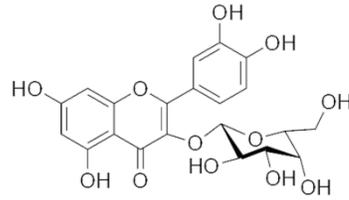 |
| keratin | $C_2H_2BrClO_2$ | 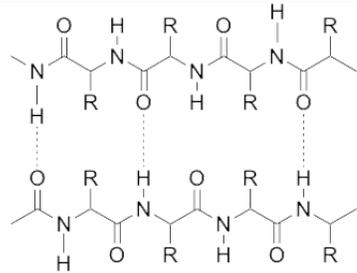 |
| zeaxanthin | $C_{40}H_{56}O_2$ | 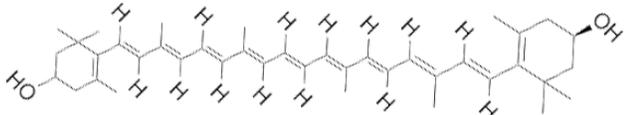 |
| melanin | $C_{18}H_{10}N_2O_4$ | 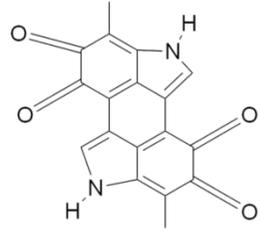 |

### 2.3.3.  Numerical Simulation of Scattering and Absorption in Biological Tissue

Monte Carlo simulations can evaluate scattering and absorption in biological tissue based on its optical properties.[51, 52] Using stochastic sampling over the trajectories of many light pockets, these simulations visually illustrate light propagation in tissues with different optical properties and can be verified by quantitative comparison between experimental results, such as diffuse reflectance from the tissue.[49] Utilizing three-dimensional simulation, we illustrated two extreme cases: absorption-dominated and scattering-dominated tissues.[53] In absorption-dominated tissues (**Fig. 2c**), light energy is rapidly absorbed, significantly decreasing penetration depth without changing the original direction. In scattering-dominated tissues (**Fig. 2d**), light exits in multiple directions, resulting in decay of light intensity along propagation.

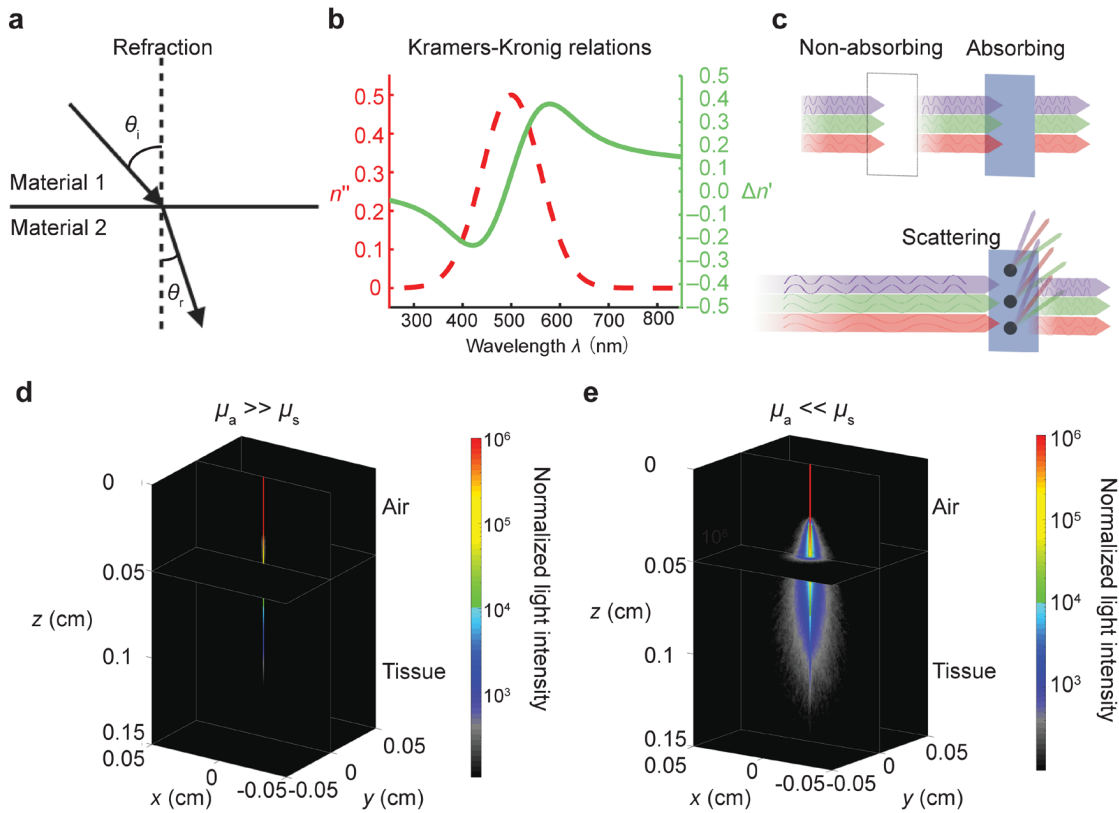

**Figure 2**. Scattering and absorption of light inside the materials. (**a**) Schematic illustration of the refraction of light when crossing an interface of two materials with different optical properties. (**b**) Numerical simulation of Kramers-Kronig relations connecting the real part of refractive index ($\Delta n' = n' - 1$) and imaginary part of refractive index $n''$. (**c**) Schematic illustration of light with different wavelengths propagating in non-absorbing, absorbing, and scattering medium. (**d-e**) Monte Carlo simulations of light propagating in hypothetical materials when absorption (**d**) and scattering (**e**) dominant, respectively.

## 3. REFRACTIVE INDEX ENGINEERING IN BIOLOGY

In this section, we discuss several unique designs of refractive index profiles found in nature to achieve different optical functionalities. Specifically, we focus on the functions of these examples in applications such as light focusing, waveguiding, optical filtering, and optical coating (**Fig. 1**).

### 3.1. Optical Focusing

One of the most important examples of refractive index design is found in the eyes of animals and humans.[5, 9] The eye is a unique organ that focuses light onto the retina to form an image, based on principles similar to those of artificial imaging systems (**Fig. 3a**).[6] To achieve this function, the eye has a distinct refractive index profile across different tissues (Fig. 3b).[5] Specifically, the lens has the highest refractive index of 1.41, while other components have relatively lower refractive indices (1.37 for the cornea, 1.33 for the aqueous and vitreous humor). The lens itself has a non-uniform refractive index, with the central refractive index being the highest. Measurements of the refractive index along the equatorial and optical axes of the lens show a pronounced curvature, peaking at the center and generally decreasing towards the periphery (**Fig. 3c**).[54] As individuals age, the refractive index profiles become flatter in the central region and steeper in the periphery. This change is an important mechanism underlying the observed alterations in the power and longitudinal aberration of the human lens.[6] Improved understanding of the refractive index distribution in the lens has applications in clinical imaging, such as optical coherence tomography, by providing more accurate measurements of the human eye.[55, 56]

### 3.2. Optical Fiber

Another example of optical design in biological systems is the waveguiding cells within the vision system (**Fig. 3d**). Among the various cells in the eye, Müller cells in the retina function like optical fibers, efficiently transmitting light with minimal dispersion from the retinal surface to the photoreceptor cell layer (**Fig. 1a**).[25] In classical optical fibers, light is confined transversely by an elevated refractive index of the core compared to its cladding, resulting in total internal reflection as discussed in Section 2.[34] Müller cell shows a similar refractive index profile to the optical fiber for light delivery (**Fig. 3e**): The mean refractive index of Müller cell stalks is 1.380, which the somata of various neurons, such as ganglion, amacrine and bipolar cells, displays a lower refractive index of 1.358. Towards the endfoot of the Müller cell, where the funnel-shaped terminal faces the vitreous body, the refractive index reduces to 1.359, reducing reflection at the interface between the vitreous and the retina (**Fig. 3e**).

Beyond the structural similarity between the Müller cell and optical fibers, the functionality of Müller cells can be quantitatively demonstrated using an optical fiber setup. The propagation of light through individual living Müller cells was assessed using a fiber-optic dual-beam laser trap (**Fig. 3f**).[25] Alongside the infrared trapping laser beams, visible light at wavelength of 514 nm was introduced into one end of the cell, and the light power transmitted to the opposite side was quantified. This setup can directly assess the axial light transmission through individual cells, as the light reentering the output fiber depends on the distance from the input fiber and the optical properties of the trapped object. With a Müller cell present in the trap surrounded by media with refractive indices up to 1.36 to mimic the retinal environment, light transmission into the output fiber was comparable to the situation where both fibers were in contact, as long as the light propagation direction was the same as in the retina. This design indicates that a single cell can act as an optical fiber in the visual system and potentially be used as waveguides inside the animal body for light delivery.

### 3.3. Optical Filter

Optical filters in biological systems can enhance the sensitivity of optical systems, especially in the vision systems (**Fig. 3g**).[57] Birds have a visual system far superior than human, and the four types of cone photoreceptor cells in birds' eyes facilitate a tetrachromatic color vision system that spans a broad spectrum of wavelengths, from ultraviolet to red.[58] Each cone photoreceptor possesses a distinct visual pigment and contains a small optical structure known as the oil droplet.[26] These oil droplets are differently colored spheres located in the inner segment of photoreceptors, directly in front of the outer segment where different absorbing molecules is concentrated inside the lipid environment (**Fig. 1a**, **Table 1**). Different molecules show distinct absorbing properties. From example, galloxanthin absorbs in the UV and blue wavelength range, 1-carotene while astaxanthin, absorbing at green/orange wavelengths. These pigmented droplet have two functions: absorbing short-wavelength light and strongly focusing longer wavelengths (**Fig. 3h**).[26] A numerical simulation examined light propagation through a tiny sphere containing astaxanthin, and oil droplet containing astaxanthin completely obstructs 500 nm light and effectively focuses 600 nm light. The modeling confirms that the refractive index of the oil droplet becomes higher due to the existence of the absorbing molecules following Kramers-Kronig relations, and this significantly increases its lens focusing power, concentrating the light wave within a few micrometers of the oil droplet, near the entrance of the cone's outer segment.

Apart from enhancing focusing capabilities, these oil droplets act as spectral filters to shift the sensitivity peaks of the photoreceptors to wavelengths longer than the peak absorption wavelengths of the pigments (**Fig. 3i**). Thus, the actual spectral sensitivity of the visual systems is greatly affected by the oil droplet acting as a spectral filter, in addition to the original absorption spectrum of the visual pigment of the individual cone photoreceptors.

### 3.4. Optical Coating

Optical coatings in biological systems, based on refractive index engineering, represent a fascinating application of natural design principles (**Fig. 3j**).[29] Here, we emphasize changes in coating properties leveraging the correlation between imaginary and real part of refractive index rather than periodic structural design in photonic crystals. One of the most noticeable example is the changes in the molecular composition of plan skin result in alterations of their color as well their scattering properties, such as an apple's skin as it ripens for light protection.[27] The background color transitions from green to yellow, and the red blush intensifies as the variety changes. Different pigments contribute to the overall color, each absorbing sunlight in different wavelength ranges (**Table 1**, **Fig. 3k**). In vacuoles of the cell, anthocyanins and other pigments form complexes known as anthocyanin vacuolar inclusions (AVIs) and the reduced scattering coefficient $\mu_s'$ and absorption coefficient $\mu_a$ can be accurately modelled in red and green apple skin based on the complex refractive index and a polydisperse Mie theory.[27] The scattering and absorption peak has a shift in the wavelength, which originates from the Kramers-Kronig relations in the refractive index as discussed in Section 2. Experimentally, the scattering and absorption properties of the plan skin can also be quantitatively followed utilizing integrating sphere (**Fig. 3l**), and by identifying the absorption and scattering fingerprints from the scattering spectra, we can infer the molecular and structural changes in the skin of the apple. This understanding allows us to appreciate how nature uses refractive index engineering to achieve functional optical coatings that adapt to environmental changes and developmental stages.

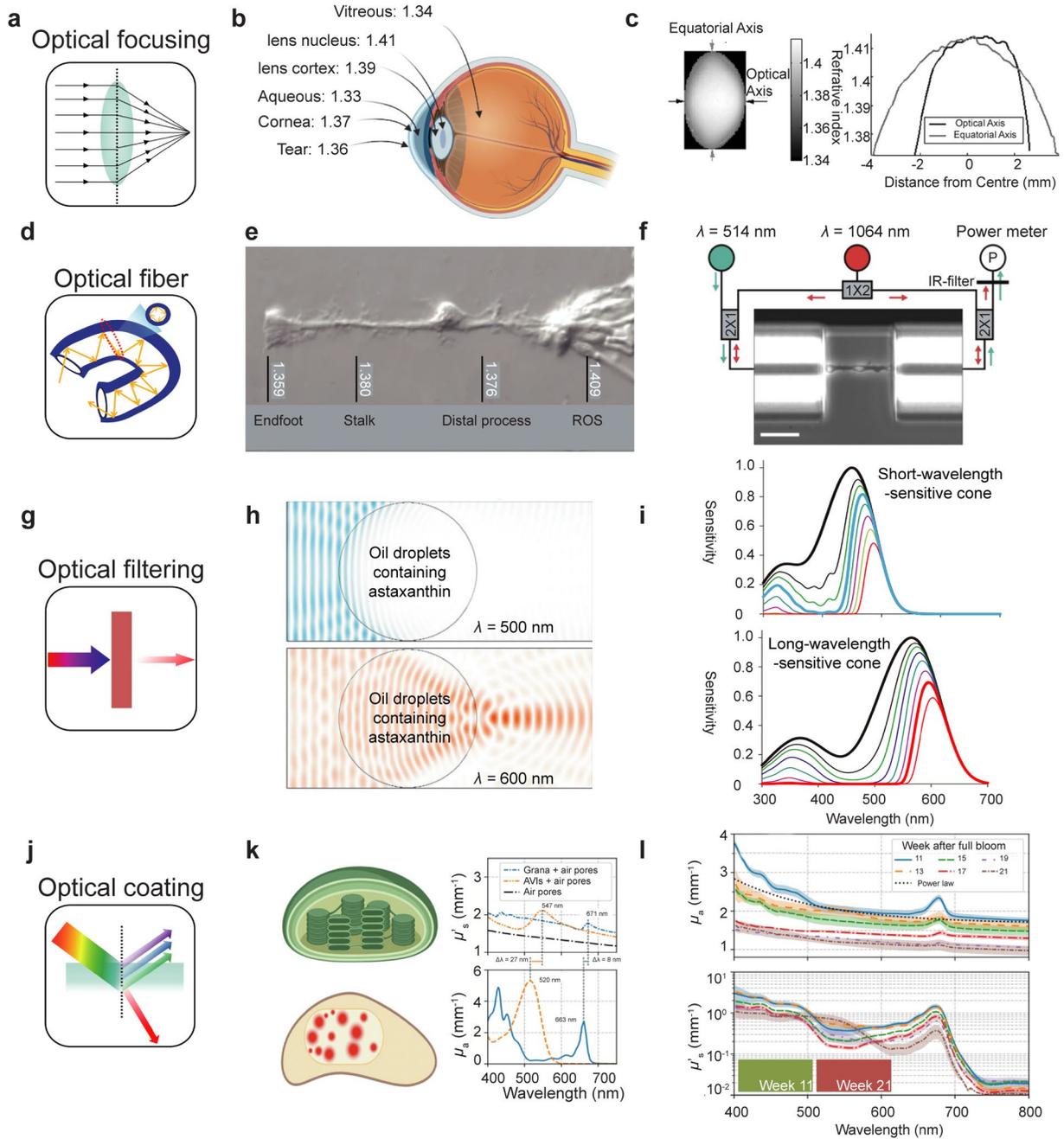

**Figure 3**. Refractive index engineering in biological systems. (**a**) Schematic of optical focusing. (**b**) Refractive index of different tissue in the eye. Reproduced with permission from ref [5]. (**c**) Refractive index map (left) and profiles along two directions (right) of the isolated human lens. Reproduced with permission from ref [54]. (**d**) Schematic of optical fiber. (**e**) Shape and refractive index of the Müller cell. (**f**) Demonstration of light guidance by an individual Müller cell. Scale bar: 50 μm. (**e-f**) Reproduced with permission from ref [25]. (**g**) Schematic of optical filtering. (**h**) Numerical modelling of light propagating through oil droplets containing astaxanthin at wavelengths of 500 (top) and 600 (bottom) nm, respectively. (**i**) The spectral sensitivity of short-wavelength-sensitive cone (top) and long-wavelength-sensitive cone (bottom) photoreceptors on oil droplet filtering. (**h-i**) Reproduced with permission from ref [26]. (**j**) Schematic of optical coating. (**k**) Modeled reduced scattering coefficient and absorption coefficient in apple skin. (**l**) Temporal evolution of the reduced scattering coefficient and absorption coefficient in apple skin. (**k-l**) Reproduced with permission from ref [27].

### 3.5. Dynamic Tunability of Optical Properties in Biological Systems

Apart from discussing specific optical examples in vision and plants, we emphasize the unique dynamic aspects to biological systems. Actuation and adaptation of optical systems are two types of responses, and here we highlight several examples in natural systems, particularly marine organisms.[59] Cephalopods utilize a unique strategy to achieve rapid dynamic changes in skin color through the mechanical action of radial muscle fibers (**Fig. 4a-c**).[18] This allows them to blend seamlessly into their surroundings. In contrast, catfish achieve near optical transparency, with a rainbow-like color visible only in transmission due to the periodic band structure of sarcomeres within their muscles (**Fig. 4d-f**).[60] Glass frogs employ a different mechanism, hiding their bodies by storing blood cells in the liver to mitigate the color caused by the strong absorption of hemoglobin.[61]

Color changes can also occur over longer timescales due to variations in background or climate. For instance, tadpoles adapt their body color from light to dark to achieve background color matching, or vice versa, over several weeks (**Fig. 4g-h**).[62] Additionally, the skin of insects often displays vibrant colors due to photonic structures, which can vary significantly with changes in the external environment, such as transitioning from dry to wet conditions (**Fig. 4i-k**).[63] The three-dimensional photonic structures in the scale of insects can be complicated, including examples from gyroid, diamond, and I-WP minimal surfaces, which also propose new questions regarding the development and evolution of biological colors.[64, 65]

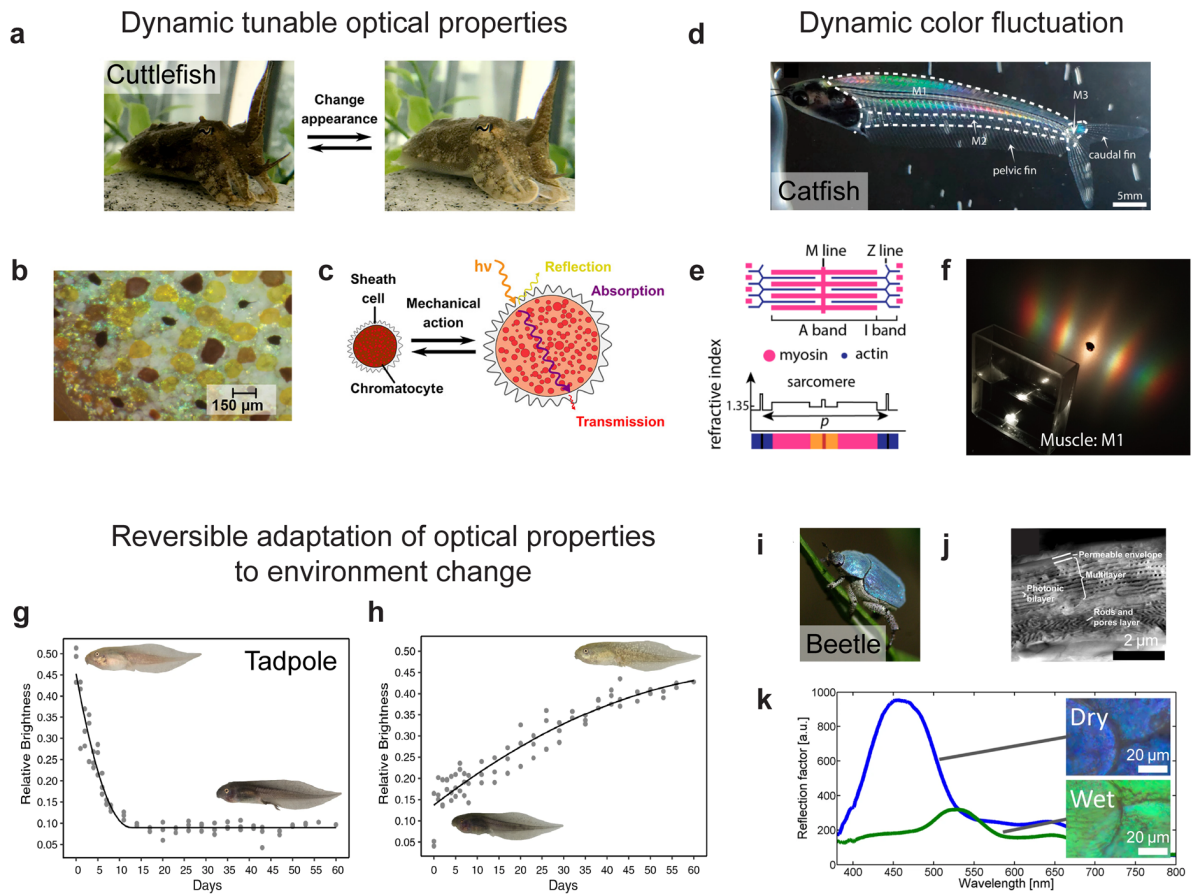

**Figure 4**. Dynamic optical properties in biological systems. (**a**) Pictures of adaptive camouflage strategy used by a cuttlefish. (**b**) Microscale images of the cuttle fish skin. (**c**) A schematic illustrates mechanical action changes the cellular organization of subcellular components, leading to change of absorption, transmission, and reflection of the skin. (**a-c**) Reproduced with permission from ref[18]. (**d**) Photograph of the catfish in natural light. (**e**) Schematic of sarcomere and subcellular refractive index profile. (**f**) White light diffraction from the fish at normal incidence. (**d-f**) Reproduced with permission from ref[60]. (**g-h**) Photos and brightness changes of tadpoles over time, for (**g**) light tadpoles in black background and (**h**) dark tadpoles in white background. of optical filtering. (**g-h**) Reproduced with permission from ref[62]. (**i**) Photograph of male *H. coerulea* beetle. (**j**) Scanning electron micrograph of the cross section of the scale. (**k**)

Reflectance changes of the scale by contacting water. (**i-k**) Reproduced with permission from ref [63, 64].

# 4. MEASUREMENT OF REFRACTIVE INDEX

In this section, we emphasize refractive index measurements for materials, especially biological tissues, relying on different optical principles. We highlight several common techniques that can be used to measure the real, imaginary, or both parts of the refractive index of materials, which are the key physical properties in this review. A more extensive discussion on the measurement of refractive index and tissue optical properties can also be found in previous reviews.[23, 24]

## 4.1. Refractometer

The refractive index of materials can be measured using a refractometer based on the total internal reflection, as shown in (**Fig. 5a**). When light passes through interfaces with different refractive indices, its propagation direction changes according to Snell's law (**Fig. 2a**).[34] In a refractometer, light is projected onto a sample with a refractive index lower than that of the prism. The light is diffused, meaning rays penetrate the sample at multiple angles, including grazing incidence. The ray at grazing incidence is refracted into the prism at the critical angle, which is the minimum angle of refraction at the prism's second face. This critical angle establishes a distinct line between light and shadow, as imaged through a telescope in the refractometer (**Fig. 5a**). Thus, the refractive index of the sample can be estimated by measuring the angle between the shadow border and the normal to the prism's second face. When different materials are loaded onto the refractometer, the maximum angle of light exiting the interface varies, showing a dark-bright interface at different locations. This procedure enables accurate assessment of the refractive index, essential for analyzing material with unknown characteristics, and refractive index at various wavelengths can be determined by changing the input light source.[41] It is noteworthy that the refractometer can also be used to measure the effective refractive index of bulk biological materials (e.g., sclera), rather than the refractive index of each subcellular component.[66] However, a comprehensive tissue model can be built to predict the optical properties of tissue based on the knowledge of the composition and structural arrangements of subcellular components, such as sclera and blood.[36]

### 4.2. Confocal Microscopy

The refractive index can also be determined using confocal microscopy due to its depth discrimination capabilities by rejecting lights that are not emitted from the focus region (**Fig. 5b**).[49] Due to the Snells law (**Fig. 2a**), the vertical translation of the focus point inside the sample is related to the geometry and refractive index of the material itself.[67] By using an illumination light containing multiple wavelengths, such as halogen light source, light of different wavelengths may focus on different locations of the sample, resulting in alterations in the optical path length due to variations in the sample's refractive index. By conducting axial scanning, the focal point will cross the sample, enabling the measurement of reflected light from different layers and different wavelengths (**Fig. 5b**). For example, distinct wavelengths $\lambda_1$ and $\lambda_2$ are associated with two focus points sites $P_1$ and $P_2$ in the imaging system. $P_1$ is located at the upper surface, indicating light doesn't go through the sample, while $P_2$ is located at the lower surface. Thus, by comparing the difference in the motor scanning distances with the actual sample thickness, the refractive index of the materials can be evaluated (**Fig. 5c**) and the multiwavelength light source also offers the capability to study the dispersion of refractive index on wavelength. Although confocal microscope allows a relatively easy setup to measure the refractive index, a comprehensive evaluation of the dependence of the refractive index on the geometry needs to be considered in the mathematical modeling and high-resolution axial scanning is necessary to acquire experimental results with sufficiently high accuracy.

### 4.3. Quantitative Phase Microscopy

The refractive index of microstructures can also be quantitatively evaluated based on the phase contrast generated from microscopy images, especially in quantitative phase microscopy. Full-field quantitative imaging methods are capable of imaging the absolute phase associated with cells and subcellular structures, relying on interferometry to measure phase delays associated with the passage of light through a given sample.[68] Technically, a coherent laser light is split into two arms, one of which interacts with the sample while the other serves as a reference beam. By interfering with the undisturbed incident wave with the transmitted wave, which is modified by the sample, quantitative images of the phase delay induced by the sample are created (**Fig. 5d**).[69] In many examples where the refractive index of the sample is already known, the resulting structural information about the sample can be reconstructed with nanometer sensitivity.[70] It is noteworthy that these structural measurements based on phase delay are not limited by the diffraction limits, which is usually the resolution limit in classical optical imaging systems.[34] As an example, this method can be used to measure the nanoscale membrane fluctuation in red blood cells, where the phase delay through the sample is directly proportional to the thickness or height of the object (**Fig. 1e**).[71] Optical imaging has the unique advantage of fast imaging time scale and with no specific sample preparation requirement. On the other hand, if structural information of the samples is already available from other measurements, such as atomic force microscopy, the spatial mapping of the refractive index can also be derived based on the two-dimensional phase image.[24, 72]

### 4.4. UV-vis-NIR Spectroscopy

The imaginary part of the refractive index $n''$ can be directly measured by comparing the intensity of light after transmission (**Fig. 5f**). According to the light propagation in a uniform medium discussed in Section 2, which describes the attenuation of light as it passes through an absorbing-only medium (**Fig. 5g**), the amount of light absorbed by the medium is described by the absorption coefficient $\mu_a$, from which the imaginary part of refractive index $n''$ can be calculated by the equation $\mu_a = 4\pi n'' \lambda^{-1}$, where $\lambda$ is the vacuum light wavelength. In UV-vis-NIR spectrometer, the instrument usually reports an absorbance of the sample defined as $Abs = -\log_{10}(I_{out}/I_{in}) = 2 - \log_{10}(T)$, where $I_{out}$ and $I_{in}$ are the output and input light intensity, respectively, and $T$ is the transmittance of light in unit of percentage. Comparing the definition of $\mu_a$ and $Abs$, the imaginary part of refractive index $n''$ can be calculated using the following relation: $n'' = -\ln(10)Abs/(4\pi d\lambda^{-1})$, where $d$ is the thickness of the sample. Compared to other methods, absorption measurements can achieve very high sensitivity, making them suitable for measuring the optical properties of low-absorbing materials by increasing the sample thickness. It is worth mentioning that although UV-vis-NIR spectroscopy can only provide direct measurements about the imaginary part of the refractive index, the real part of the refractive index can potentially be inferred from the absorption spectrum based on the Kramers-Kronig relations discussed in Section 2 (**Fig. 2b**). This connection demonstrates how changes in the refractive index, especially in its imaginary portion, can be utilized to quantitatively modulate the optical properties of materials, especially absorption and scattering in the tissues.

Another application of UV-vis-NIR spectroscopy is to measure the refractive index of unknown scatterers, such as cells and nanoparticles, by varying the background solution refractive index $n'_{background}$ (**Fig. 5g**). By maximizing the transmittance of light through a scattering suspension, the refractive index of the scatterer $n'_{scatterer}$ can be estimated (**Fig. 5h**). However, this method can only provide an effective refractive index for complex scatterer, such as cells, without information about spatial heterogeneity.[23, 24] Combining the UV-vis-NIR spectrometer with an accessory of integrating sphere, the absorption coefficient $\mu_a$ and scattering coefficient $\mu_s$ of biological tissues can also be decoupled and measured separately based on different sample loading configurations.[45]

### 4.5. Ellipsometry

Ellipsometry is an optical method that evaluates how light changes in polarization upon reflection from single or multiple optical interfaces, enabling the determination of thin film thickness and refractive index (**Fig. 5i**).[73] The measured signal represents the alteration in polarization resulting from the interaction of incident radiation, in a defined state, with the material structure of interest, whether it be reflected, absorbed, scattered, or transmitted. The change in polarization is quantified through the amplitude ratio $\psi$ and phase difference $\Delta$. The polarization state of the incident light can be decomposed into two components depending on the polarization direction. One component oscillates perpendicular to the plane of incidence and parallel to the sample surface, while the other component oscillates parallel to the plane of incidence (**Fig. 5i**). The real and imaginary part of refractive index can be calculated from the $\psi$ and $\Delta$ data over the wavelength range (**Fig. 5j**). By setting up a mathematical model that fits the geometry of the measurement, the complex refractive index can be calculated from the experimental measurements analytically, and only solutions that satisfy physical constraints such as Kramers-Kronig relations and positive values for $n''$ are meaningful. Ellipsometry has been widely used to provide precise optical characterization of thin films as well as the thickness of materials, offering an efficient method for analyzing ultra-thin materials and multilayer structures. For a simplified geometry of single interface reflection in the air, the complex refractive index $n$ can be easily calculated (**Fig. 5k**) from measurements of $\psi$ and $\Delta$ using $n = \sqrt{\sin^2(\theta_i) \left[ 1 + \tan^2(\theta_i) \left( \frac{1-\rho}{1+\rho} \right)^2 \right]}$, where $\theta_i$ is the angle of light incident onto the interface (**Fig. 2a**) and $\rho$ is a complex dimensionless parameter defined as $\rho = \tan(\psi) \exp(i\Delta)$.

Although an ellipsometer can be used to measure refractive index accurately, it has several requirements. First, it requires a flat interface so that the reflected light can be collected and analyzed by the detector. Second, it requires pre-knowledge of the measurement geometry so that a theoretical or numerical model can be built to evaluate the optical properties of the unknown materials. Third, it has relatively low accuracy in measuring the imaginary part of refractive index in comparison to other methods, such as transmittance measurement in the UV-vis-NIR spectroscopy.

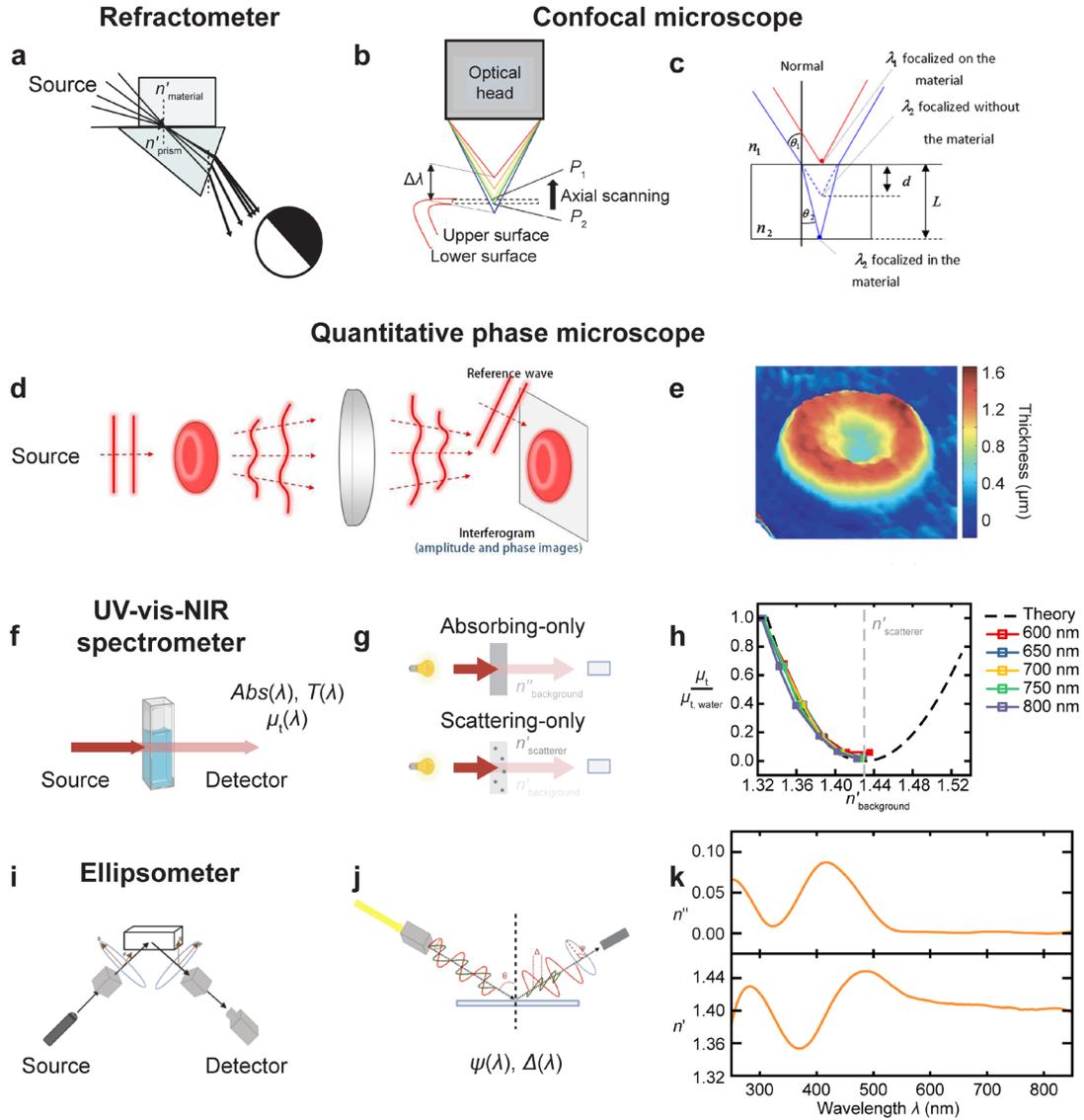

**Figure 5**. Techniques to measure the refractive index of materials. (**a**) Schematic illustrating the working principle of refractometer. (**b**) Schematic illustrating the different focus from a chromatic confocal microscope. (**c**) Different spectral components focalized inside and outside the material. Reproduced with permission from ref [67]. (**d**) Schematic illustrating the principle for quantitative phase imaging. Reproduced with permission from ref [69]. (**e**) Two-dimensional thickness profile of a living red blood cell captured by quantitative phase imaging. Reproduced with permission from ref [71]. (**f**) Schematic illustrating the working principle of UV-vis-NIR

spectrometer. (**g**) Schematic illustrating the measurement for absorbing and scattering samples utilizing UV-vis-NIR spectrometer. (**h**) Dependence of scattering coefficient in a scattering phantom containing silica particles on the refractive index of the background. (**i**) Schematic illustrating the working principle of ellipsometer. (**j**) Raw data extracted from ellipsometer. (**k**) Refractive index (bottom: real, top: imaginary) exacted from the ellipsometer using the simplified single-interface geometry model. (**h, k**) Reproduced with permission from ref [74].

### 4.6. Summary of Different Techniques for Measuring Refractive Index

Measurement of the refractive index can be broadly divided into two categories: bulk measurements and microscopic measurements. Bulk measurements are used for samples with uniform refractive index values (**Fig. 6a**), while microscopic measurements quantify refractive index variations at the microscale (**Fig. 6b**). In many practical applications, microscale components are purified, and their optical properties are measured using bulk techniques.[23, 24] However, the ability to measure microscopic refractive index variations without damaging the sample offers unique insights into the sample's structure and function. This capability is particularly valuable for noninvasive biomedical applications, such as optical coherence tomography (OCT), where detailed information about refractive index variations can enhance imaging depth and resolution.[56] High-resolution imaging relying on understanding of the refractive index of tissue can also dynamically imagine the fluctuation in living cells.[71] By leveraging these advanced measurement techniques, a deeper understanding of light interaction with biological tissues can be achieved for developing innovative solutions for medical diagnostics and treatment.[3, 37]

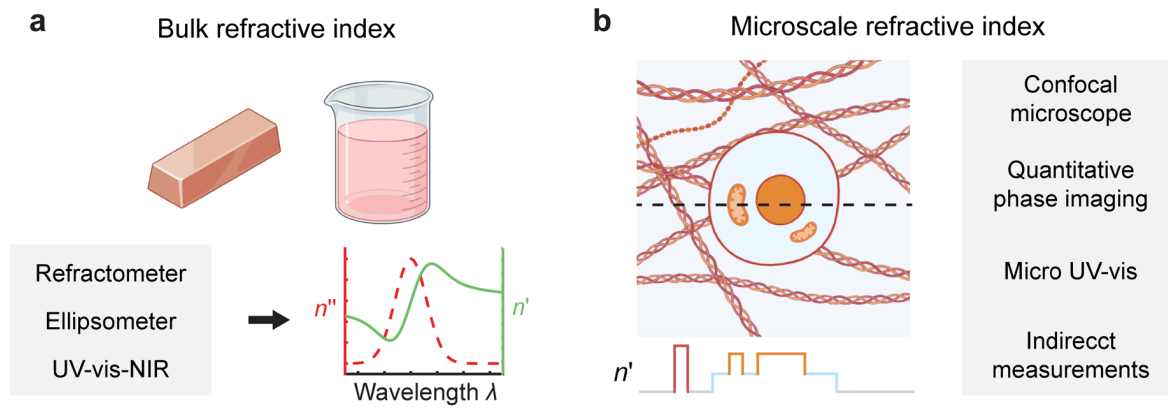

**Figure 6**. Comparison of techniques to quantify bulk and microscale refractive index properties. (**a**) Techniques from which refractive index of bulk refractive index can be measured. (**b**) Techniques used to quantify the spatial heterogeneity of refractive index in biological samples. The colored line at the bottom highlights the spatial variation of real refractive index $n'$ along the dashed line across the biological tissue.

Comparison of various techniques for measuring the refractive index of materials are provided in **Table 2**, each with distinct advantages and limitations. Refractometers are valuable for directly measuring the real part of the refractive index through total internal reflection, requiring minimal sample amounts. However, they are limited to bulk characteristics and are unsuitable for heterogeneous samples although an average refractive index of thin tissue slice can be measured.[75] UV-Vis-NIR spectrometers provide high-accuracy in measuring the imaginary part of the refractive index, yet they also primarily assess bulk properties. Ellipsometers offer the capability to measure both real and imaginary parts of the refractive index and are highly sensitive to thin optical samples, including 2D materials. Nonetheless, they necessitate a smooth optical interface for effective reflection measurements.

Confocal microscopes provide high spatial resolution, constrained by the diffraction limit, and high temporal resolution, making them ideal for dynamic analysis. However, they are sensitive to noise, such as autofluorescence. Quantitative phase imaging similarly offers high spatial and temporal resolution but requires detailed information about sample height profiles.[72] Micro UV-vis spectrometers are advantageous for their high spatial resolution but are limited to measuring the imaginary part of the refractive index.[76] Apart from the methods mentioned above, other indirect measurement techniques has been proposed for mapping refractive index with subcellular spatial resolution and high temporal resolution, suitable for dynamic analysis, though careful evaluation of data interpretation is essential.[47, 48]

**Table 2. Comparison of different techniques for measuring refractive index of materials.**

| Technique | Advantages | Limitations |
|---|---|---|
| Refractometer | - Direct measurement of real part of refractive index based on total internal reflection<br>- Small sample amount required | - Only accessible to real part of refractive index<br>- Typically, it assesses bulk characteristics.<br>- Cannot be applied for heterogeneous samples |
| UV-Vis-NIR spectrometer | - Direct measurement of imaginary part of refractive index<br>- High measurement accuracy | - Only accessible to imaginary part of refractive index<br>- Typically, it assesses bulk characteristics. |
| Ellipsometer | - Accessible to both real and imaginary part of refractive index<br>- Sensitive to thin optical samples, even 2D materials | - Typically, it assesses bulk characteristics.<br>- Sample needs to have a smooth optical interface for reflection |
| Confocal microscope | - High spatial resolution limited by diffraction limit<br>- High temporal resolution for potential dynamic analysis | - Sensitive to noise like autofluorescence |

| Quantitative phase imaging | - High spatial resolution limited by diffraction limit<br>- High temporal resolution for potential dynamic analysis | - Require information about sample height profiles |
| Micro UV-vis spectrometer | - High spatial resolution limited by diffraction limit | - Only accessible to imaginary part of refractive index |
| Indirect measurements | - High spatial resolution, e.g., nanometer scale<br>- High temporal resolution for potential dynamic analysis | - Interpretation of data needs to be carefully evaluated |

## 5. DISCUSSION

### 5.1. Biomimetic Materials Design

Understanding the physical principles behind biological structures is crucial for the design of innovative biomimetic materials.[11, 13, 20, 77-79] Biological systems have evolved over millions of years to optimize their structures for specific functions, offering a rich source of inspiration for material scientists.[8] From the optical perspective, by studying these natural systems, we can identify the molecular building blocks and structural motifs that contribute to their unique optical properties. One unique feature in biological system is the similar molecular composition of different types of tissues but the change of portion of different components, they can achieve distinct optical properties. In the example of eye (**Fig. 7a**), the high refractive index profiles in the lens are achieved by the relatively high fraction of proteins, lipids, and ions, etc.[5] Similarly, the relatively low refractive index in the vitreous results from the high fraction of water inside. By leveraging these principles, optical systems can be designed with similar molecular compositions but tailored properties for each unit, such as varying refractive indexes, mechanical strength, and biocompatibility.

## 5.2. Intricate Optical Systems Based on Refractive Index Contrast

Optical systems in nature often rely on subtle variations in refractive index to achieve complex functions. For example, the Muller cells in the eye use subtle changes in the refractive index of cells to guide light efficiently, acting as optical fibers to improve light sensitivity in the eye (**Fig. 7b**).[25] In artificial optical systems, optical fibers are a practical example of how refractive index contrasts can be exploited to guide light over long distances with minimal loss.[34] These fibers usually consist of a core with a slightly higher refractive index than the surrounding cladding, creating total internal reflection that confines the light within the core. This principle can be extended to design intricate optical devices, such as waveguides, photonic crystals, and metamaterials, which manipulate light in novel ways.[80] By carefully engineering the refractive index profile, we can create devices with enhanced performance for applications in telecommunications, imaging, and sensing.

## 5.3. Optical Scattering Engineering

The manipulation of optical scattering properties through refractive index engineering holds significant potential for biomedical imaging and other applications.[3] The refractive index of a material determines how much light is scattered and absorbed as it passes through. Apart from the transparent embryo of zebrafish that has been broadly used in developmental studies, nature has provided strategies in glassfrogs[61] and ghost catfish[60] where changing in transparency can be achieved in adults. Inspired by the promise of creating transparent animals, a new strategy to tune the scattering and absorption characteristics of tissues was developed based on changing the real and imaginary parts of the refractive index for live mice (**Fig. 7c**).[74, 81] This is particularly important in biomedical imaging, where strong scattering of visible light by tissues limits the depth and resolution of imaging techniques such as optical coherence tomography (OCT)，photoacoustic imaging, and multiphoton microscopy.[82-84] By developing materials with tailored scattering properties, we can enhance the penetration depth and contrast of existing imaging modalities.[85-88] Additionally, this approach can be used to design optical phantoms that mimic the scattering properties of biological tissues, providing valuable tools for the calibration and validation of imaging systems.[49, 89] Beyond biomedical applications, optical scattering engineering can improve the performance of devices in areas such as photovoltaics, where controlling light absorption and scattering can enhance the efficiency of functional devices such as solar cells.[90, 91]

### 5.4. Biomimetic Optical Devices

Beyond the intricate design of optical properties through microscopic engineering of the refractive index, the propagation of light through biological systems can directly inspire practical applications. For instance, in the visual system, mimicking the human retina (as shown in **Fig. 1a**), a high-density array ($>10^9$ cm$^{-2}$) of nanowires is proposed to replicate the dense photoreceptors in the eye (**Fig. 7d**).[92] Metal oxides enable synaptic bidirectional photo-response, while the Pb electrode at the bottom absorbs additional light to prevent aberration. The generated photocurrent exhibits different waveforms for various colors, facilitating filter-free color vision. [92] Additionally, the gradual increase and decrease of current in response to light intensity mimics the behavior of light in actual neural activities.

In another example, designing a reflecting interface inspired by the camouflage strategies of cephalopods (as shown in **Fig. 4a-c**), a hierarchical structure combining copper nanostructures, conductive co-polymer, and acrylic dielectric elastomer has been proposed (**Fig.**

**7e**). The device's optical properties and visual appearance change under different mechanical stresses. By altering the strain from positive (compressed) to negative (extended), the device's reflectance and transmittance gradually increase while absorption decreases. This allows the device to modulate its optical properties and visual appearance based on mechanical actuation.

**a**  Biomimetic materials design

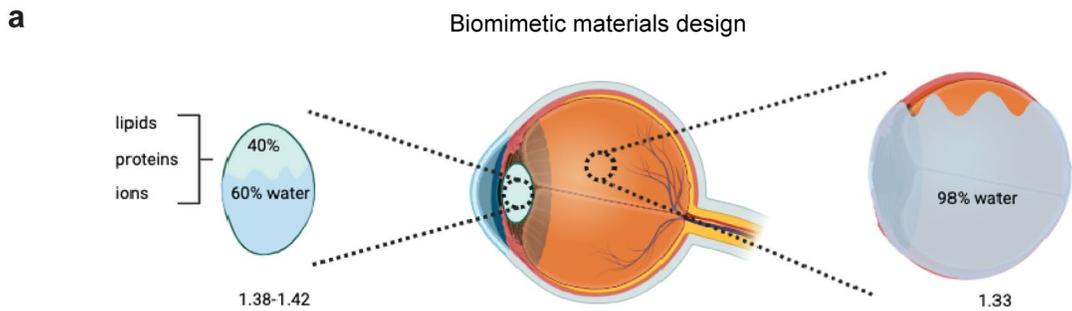

**b**  Intricate optical systems based on refractive index contrast

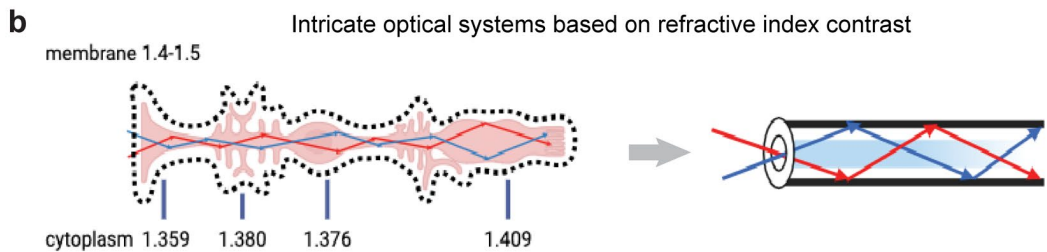

**c**  Optical scattering engineering for biomedical imaging

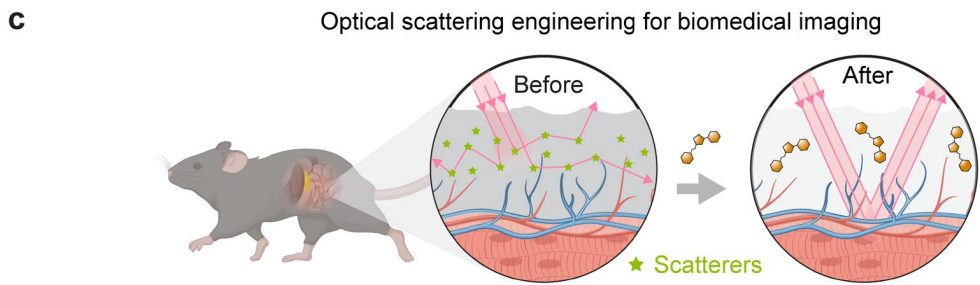

**d**  Biomimetic design of imaging systems

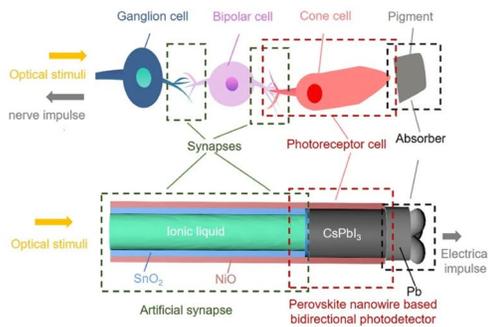

**e**  Biomimetic design of reconfigurable reflectance

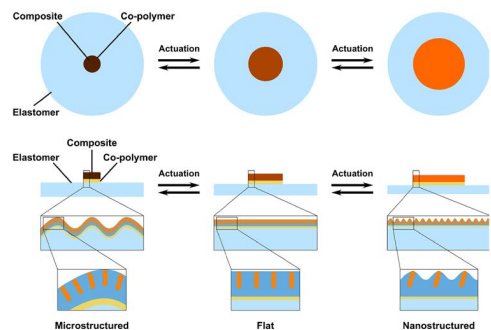

**Figure 7**. Refractive index engineering inspired by natural systems. (**a**) Biomimetic materials design of different refractive index materials based on tunning the composition level of similar molecular structures. (**b**) Intricate optical systems based on small variations in the optical properties. (**c**) Modulating the scattering properties of tissue by changing the level of absorbing molecules inside. Reproduced with permission from ref [74]. (**d**) Schematics of an artificial neuromorphic eye based on nanowire structure mimicking human retina. Reproduced with permission from ref[92]. (**e**) Schematic of a cephalopod-inspired surface, where mechanic actuation can leads to change of surface reflectance, transmittance, and absorption. Reproduced with permission from ref [74]. Copyright 2021 American Chemical Society.

## 5.5. Theoretical and Numerical Tools for Refractive Index Engineering

The advancement of theoretical modeling and numerical tools, particularly machine learning, has provided new strategies for engineering the refractive index of materials. While we discussed the connection between absorption and refractive index in Section 2, the physical origin of absorption of materials are not discussed and the optical properties of molecules or materials can also be modeled using polarizability models and calculated using first-principle density functional theory (DFT) calculations.[93, 94] These models allow for a detailed understanding of how light interacts with materials by correlating the absorption coefficient of realistic dye molecules with accurate polarizability models.[93] An accurate understanding of the refractive index is crucial for predicting and optimizing the light-matter interaction at the nanoscale.[95]

In addition to first-principle theoretical methods, the development of machine-learning algorithms has revolutionized materials discovery, particularly for refractive index engineering.[96]

By leveraging existing optical databases, especially those related to the optical absorption properties of chemicals, machine learning can predict the absorption properties of unknown molecules, significantly improving the efficiency of identifying new materials.[91, 97] For example, machine-learning models using structures such as random forests or support vector machines can screen materials for specific absorption properties within certain wavelength ranges with an accuracy of approximately 90%.[98] A critical step in machine learning is identifying accurate molecular descriptors for input structures and machine learning algorithm can also be combined with density functional theory to select optimized molecular features, providing accurate predictions of the optical properties of large molecules, including oligomers.[99]

## 5.6. Challenges and Perspectives on Refractive Index Engineering

One of the significant challenges in refractive index engineering is the precise control of optical properties at very small scales, including the subcellular level.[23] Biological tissues exhibit complex refractive index profiles due to the heterogeneous nature of their subcellular components, as we have discussed in Section 2. Achieving accurate control over these properties requires advanced techniques that can manipulate the refractive index at the nanoscale. This involves not only understanding the intrinsic optical properties of individual components but also how these components interact and assemble with each other to influence the overall optical behavior of the tissue. Although top-down fabrication has been the major focus in industry application, nature has been adapting the strategy of bottom-up self-assembly, where all different biological structures share similar building blocks.[100, 101] Future advancements in nanotechnology and material science will be crucial in developing methods to engineer these properties with high precision, enabling new applications in biomedical imaging and diagnostics.

Another frontier in refractive index engineering is the ability to dynamically change the optical properties of materials in response to external stimuli (**Fig. 8a**). Nature provides numerous examples of dynamic optical systems, such as the color-changing skin of cephalopods[14, 18] and the adaptive lenses in the eyes of certain animals.[6] Mimicking these dynamic systems in synthetic materials requires the development of responsive materials that can alter their refractive index in real-time. This could be achieved through the incorporation of stimuli-responsive polymers, liquid crystals, or other smart materials that react to changes in temperature, light, electric fields, or mechanical stress.[18] The ability to dynamically tune optical properties opens up possibilities for advanced applications in adaptive optics, smart windows, and responsive sensors, where materials can adjust their behavior based on environmental conditions.[92]

The morphological evolution of structures, as observed in nature, presents both a challenge and an opportunity for refractive index engineering (**Fig. 8b**). Natural systems often exhibit hierarchical and adaptive structures that evolve over time to optimize their optical functions.[78, 101] For example, the layered structure of butterfly wings or the photonic crystals in peacock feathers are results of evolutionary processes that enhance their optical performance. [29, 63, 64] Replicating these complex and evolving structures in synthetic materials requires a deep understanding of hierarchical self-organization. Advances in materials science, particularly in the fields of biomimetics and nanofabrication, are essential for creating materials that can mimic these natural processes. By harnessing the principles of morphological evolution, we can develop materials with enhanced optical properties that can adapt to changing environments, much like their natural counterparts.

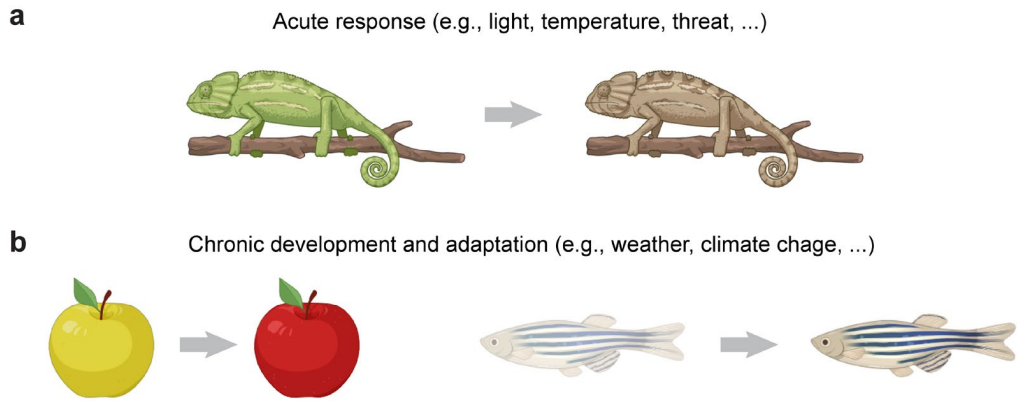

**Figure 8**. Response and adaptation of optical properties in biological systems. (**a**) Acute response of skin color to external environment (e.g., chameleons). (**b**) Chronic adaptation of skin colors to climate change or weather (e.g., apple skin, fish transparency).

## 6. CONCLUSIONS

In summary, the design of biomimetic materials, intricate optical systems, and optical scattering engineering based on refractive index manipulation offers exciting opportunities for advancing technology in various fields. By drawing inspiration from nature and understanding the underlying physical principles, we can develop materials and devices with enhanced performance and new functionalities. We also highlight the importance of optical scattering engineering in biomedical imaging, where controlling light scattering and absorption can significantly improve imaging depth and resolution. Techniques such as refractometry, confocal microscopy, quantitative phase microscopy, UV-vis-NIR spectroscopy, and ellipsometry provide comprehensive insights into the optical properties of biological materials, facilitating the development of advanced biomimetic designs. Future research should focus on further exploring these principles and translating them into practical applications to address real-world challenges. By continuing to draw inspiration from nature and leveraging our understanding of the physical principles behind biological structures, we can create innovative materials and devices that push the boundaries of current technology. Moreover, interdisciplinary collaborations between materials science, physics, and biomedical engineering can play a crucial role in developing innovative optical materials. These fields can work together to harness the unique properties of biological systems and translate them into practical applications. The implications of refractive index engineering in emerging technologies, such as advanced imaging systems or bio-integrated optical devices, offer a promising direction for future exploration. By fostering such collaborations, the development of cutting-edge technologies can be accelerated to address complex challenges and enhance our understanding of the natural world.

AUTHOR INFORMATION

**Corresponding Author**

Zihao Ou - *Department of Physics, The University of Texas at Dallas, Richardson, TX, 75080, United States*; https://orcid.org/0000-0003-2987-7423; Email: Zihao.Ou@UTDallas.edu

**Acknowledgements**

Schematics were prepared using BioRender.

**Author Contributions**

The manuscript was written through contributions of all authors. All authors have given approval to the final version of the manuscript.

**Funding Sources**

This work was supported by internal funding from the University of Texas at Dallas.